\documentclass[12pt,prl,aps,amssymb,amsmath,tightenlines]{revtex4} 

\usepackage[dvips]{graphicx}
\usepackage{dsfont}

\newcommand{\re}{{\rm e}}
\newcommand{\ri}{{\rm i}}
\newcommand{\rd}{{\rm d}}

\usepackage{xcolor}

\begin{document}
\title{Biorthogonal systems on unit interval and \\ zeta dilation operators}
\author{Dorje C. Brody}

\affiliation{
Department of Optical Physics and Modern Natural Science, St Petersburg
National Research University of Information Technologies, Mechanics and Optics, 
St Petersburg 197101, Russia}

\begin{abstract}
An elementary `quantum-mechanical' derivation of the conditions for a system of functions to form a Reisz basis of a Hilbert space on a finite interval is presented.  
\end{abstract}

\maketitle

Consider a Hilbert space ${\cal H}=L^2[(0,1),\rd x]$ of square-integrable functions over 
the unit interval $(0,1)$. We are interested in the set of basis states for ${\cal H}$. It 
suffices to consider, say, odd functions. Then the only set of orthonormal basis functions 
is given by $\{e_n(x)\}$ with 
\[
e_n(x) = \sqrt{2} \sin(n\pi x) . 
\] 
For many applications, however, all we require from a basis set $\{\varphi_n(x)\}$ is that 
it should be complete. The orthogonality condition satisfied by $\{e_n(x)\}$ for sure often 
simplifies calculations in practice, but this is not essential. Indeed, a closed quantum system 
modelled by a non-Hermitian Hamiltonian, for instance the Hamiltonian of a PT-symmetric 
system, has a complete set of eigenstates that are not orthogonal (with respect to the 
Lebesgue measure). It is therefore of interest 
to examine conditions required for a basis set $\{\varphi_n(x)\}$ to be complete. When the 
Hilbert space consists of the second Lebesgue class $L^2[(0,1),\rd x]$ of square-integrable 
functions on a finite interval (such as the state space of a particle trapped in a potential well), 
it turns out that the relevant conditions on $\{\varphi_n(x)\}$ involve analytic properties of 
generalised zeta functions on the complex plane. The purpose of the present note is to 
illustrate this result using elementary quantum-mechanical techniques. 

The question to be addressed can be stated more precisely as follows. Let $\varphi(x) \in 
{\cal H}$, where ${\cal H}=L^2[(0,1),\rd x]$. What is the condition on $\varphi(x)$ such that 
the set of functions 
\[ 
\left\{ \varphi(x), \varphi(2x), \varphi(3x), \varphi(4x), \cdots \right\}
\] 
forms a complete basis? This question was addressed by Wintner (1944) who examined 
a special class of functions that can be expressed in the parametric form 
\[ 
\varphi(x) = \sum_{n=1}^\infty \frac{e_n(x)}{n^s} ,
\] 
and identified the relevant condition on this class. An analogous question was 
discussed in a seminar by Beurling (1945), and investigated further by Bourgin (1946) and 
by Neuwirth \textit{et al}. (1970). More recently, a complete characterisation of the 
solution to the problem was obtained in an influential paper by Hedenmalm \textit{et al}. 
(1997), which can be summarised briefly as follows. Because $\varphi(x)\in{\cal H}$, it 
admits an expansion 
\[ 
\varphi(x) =  \sum_{n=1}^\infty a_n \, e_n(x)
\] 
in terms of $e_n(x) = \sqrt{2} \sin(n\pi x)$. Given the set $\{a_n\}$ of coefficients, 
consider the Dirichlet series 
\[ 
L_a(s) = \sum_{n=1}^\infty \frac{a_n}{n^s} . 
\] 
Then the system $\{\varphi(nx)\}$ is a Riesz basis in ${\cal H}$ if and only if $L_a(s)$ 
is an analytic function bounded away from zero and infinity in the half plane $\Re(s)>0$. 
The appearance of the generalised zeta function $L_a(s)$ in this context might at first 
seem surprising, however, treating the problem in a purely quantum-mechanical formalism 
we see how it arises quite naturally. 

In general, when dealing with a complete set of basis 
elements $\{\varphi_n(x)\}$ that lack orthogonality, 
one is looking for its biorthogonal counterpart 
$\{{\tilde\varphi}_n(x)\}$ (see Brody 2014). Specifically, 
suppose that $(\{\varphi_n(x)\},\{{\tilde\varphi}_n(x)\})$ is a complete biorthonormal set 
of bases in ${\mathcal H}$. Then the set 
$\{\varphi_n(x)\}$ is called a `Fischer-Riesz' basis if (a) for any $\psi(x)\in{\mathcal H}$ 
we have 
\[ 
\sum_{n=1}^\infty \left| \int_0^1 \overline{{\tilde\varphi}_n(x)} \psi(x) \rd x \right|^2 <\infty; 
\] 
and (b) if for any sequence $\{c_n\}$ such that $\sum_n|c_n|^2<\infty$ there exists a 
$\psi(x)\in{\mathcal H}$ for which 
\[ 
\int_0^1 \overline{{\tilde\varphi}_n(x)} \psi(x) \rd x=c_n. 
\] 
A theorem of Bari (1951) then shows that: 
(i) $\{{\tilde\varphi}_n(x)\}$ is a Fischer-Riesz basis if and only if there exists a bounded invertible 
linear operator ${\hat u}^{-1}$ and a complete orthonormal basis elements $\{e_n(x)\}$ in 
${\mathcal H}$ such that ${\hat u}^{-1}\varphi_n(x)=e_n(x)$; and that (ii) $\{\varphi_n(x)\}$ 
is a Fischer-Riesz basis if and only if there exists a positive bounded invertible linear operator 
${\hat g}^{-1}$ in ${\mathcal H}$ such that $\varphi_n(x)={\hat g}^{-1}{\tilde\varphi}_n(x)$, where 
${\hat g}=({\hat u}{\hat u}^\dagger)^{-1}$. 

We require one further ingredient before we can proceed, namely, a quantum-mechanical 
characterisation of dilation, which has been used effectively in Bender \& Brody (2018). 
That is, the generator of dilation is given by the operator ${\hat x}{\hat p}$, where ${\hat p} 
= -\ri \, \rd/\rd x$, so that for a smooth function $f(x)$ we have 
\[ 
\re^{{\rm i}\lambda{\hat x}{\hat p}} f(x) = f(\re^\lambda x) . 
\] 
While this result is well known, for the benefit of readers less acquainted with properties of 
dilation let us remark that by defining $q=\ln x$ for $x>0$ we find for its associated `momentum' 
operator $-\ri\,\rd/\rd q=-\ri(\rd x / \rd q)
\rd/\rd x$, but $x=\re^q$ so $-\ri\,\rd/\rd q=-\ri x \, \rd/\rd x$ (cf. Twamley \& Milburn 2006). Hence 
$\re^{{\rm i}\lambda{\hat x}{\hat p}} f(x) = \re^{\lambda {\rm d}/{\rm d}q} f(\re^q)=f(\re^{q+\lambda})$, 
but a shift in $q$ is equivalent to a scaling in $x$, from which the dilation property can be inferred. 

With these observations at hand let us examine the series expansion of $\varphi(x)$ in 
${\cal H}$: 
\[ 
\varphi(x) = \sum_{n=1}^\infty a_n \sqrt{2} \sin(n\pi x) = 
\sum_{n=1}^\infty a_n n^{{\rm i}{\hat x}{\hat p}} \sqrt{2} \sin(\pi x) = 
L_a(-\ri{\hat x}{\hat p}) \, \sqrt{2} \sin(\pi x) . 
\] 
On the other hand, we have $\varphi(nx)=n^{{\rm i}{\hat x}{\hat p}}\varphi(x)$, so commuting 
$n^{{\rm i}{\hat x}{\hat p}}$ through $L_a(-\ri{\hat x}{\hat p})$ and writing $\varphi_n(x)=
\varphi(nx)$ we deduce that 
\[ 
\varphi_n(x) = L_a(-\ri{\hat x}{\hat p}) \, e_n(x) . 
\] 
In other words, the operator $L_a(-\ri{\hat x}{\hat p})$, which might be called a zeta dilation 
operator, plays the role of ${\hat u}$ indicated 
above, except that we require $L_a(-\ri{\hat x}{\hat p})$ be bounded and invertible. Now the 
eigenvalues of the operator ${\hat x}{\hat p}$ has the form $\frac{1}{2}(E+\ri)$, where $E$ 
is real. This can be seen by observing that ${\hat x}{\hat p}$ is related to the `Berry-Keating' 
Hamiltonian ${\hat h}={\hat x}{\hat p}+{\hat p}{\hat x}$ via the relation  ${\hat x}{\hat p}=
({\hat h}+\ri)/2$. Hence the eigenvalues of $-\ri{\hat x}{\hat p}$ take the form $\frac{1}{2}
(1-\ri E)$, i.e. they have a strictly positive real part. It follows that if $L_a(s)$ is an analytic 
function bounded away from zero and infinity in the half plane $\Re(s)>0$, then ${\hat u}=
L_a(-\ri{\hat x}{\hat p})$ is bounded and invertible, fulfilling the required condition on 
$\{\varphi_n(x)\}$ to form a Fischer-Riesz basis. 

We therefore arrive at an instance where a sufficiently advanced and technical idea in 
mathematics can be grasped at an elementary and intuitive level by employing an operator 
formalism of quantum theory. As for the biorthogonal counterpart, we have 
\[ 
{\tilde\varphi}_n(x) = \left(L_{\bar a}(\ri{\hat p}{\hat x})\right)^{-1} e_n(x) . 
\] 
As noted in Hedenmalm \textit{et al}. (1997), while we have $\varphi_n(x)=\varphi(nx)$, 
its biorthogonal counterpart cannot be expressed in the form ${\tilde\varphi}(nx)$ for, 
writing $\mu(n)$ for the M\"obius function and assuming that $a_1=1$ we have 
\[
\frac{1}{L_a(s)} = \sum_{n=1}^\infty \frac{\mu(n) a_n}{n^{s}} . 
\] 

\begin{figure}[t]
\begin{center}
\includegraphics[angle=0,scale=1.0]{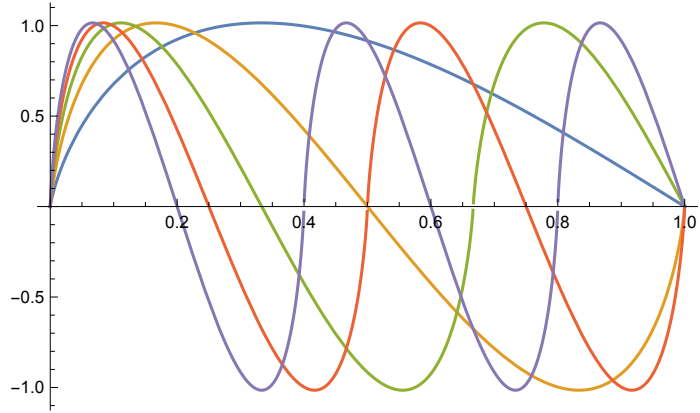}
\footnotesize
\caption{(colour online) 
\textit{The basis set $\{\varphi(nx)\}$ is plotted here for $n=1,2,\ldots,5$ when 
$\varphi(x)=\ri\left(  {\rm Li}_2(\re^{-{\rm i}\pi x})-{\rm Li}_2(\re^{{\rm i}\pi x}) \right)/\sqrt{2}$. 
These functions appear to be distorted versions of the orthonormal basis elements 
$\{\sqrt{2}\sin(n\pi x)\}$. They are not orthogonal, but they are nevertheless complete in 
that an arbitrary square-integrable function on $(0,1)$ can be expanded in terms of 
$\{\varphi(nx)\}$. 
\label{F1}}}
\end{center}
\end{figure}
 
As an illustrative example, let us consider the function 
\[ 
\varphi(x) = \frac{\ri}{\sqrt{2}}\left(  {\rm Li}_k(\re^{-{\rm i}\pi x}) - 
{\rm Li}_k(\re^{{\rm i}\pi x}) \right) , 
\] 
where ${\rm Li}_k(x)$ denotes the polylogarithm function. Then we have $a_n=n^{-k}$ 
and for $\Re(k)>\frac{1}{2}$ the set of functions $\{\varphi(nx)\}$ behaves like a set of 
distorted sine functions on $(0,1)$ that nevertheless can be used to expand any 
function in ${\cal H}$. In this case the relevant zeta dilation operator is just the Riemann 
zeta function evaluated at a shifted dilation generator: 
\[ 
\varphi(x) = 2 \, \zeta(k-\ri{\hat x}{\hat p}) \, \sin(\pi x) .
\] 
In particular, for $k=0$, which may be defined via an analytic continuation, we get 
the Riemann dilation operator $\zeta(-\ri{\hat x}{\hat p})$, whose eigenvalues are the 
Riemann zeta functions $\zeta(s)$ with $s=\frac{1}{2}(1-\ri E)$, $E$ real (Bender \& 
Brody 2018). Examples of the first five basis functions for $k=2$ are illustrated in 
Fig.~\ref{F1}. 

Finally, a curious physics student might ask whether an analogous construction exists 
for second Lebesgue class functions defined on the whole line. It suffices to consider a 
Gaussian Hilbert space. Here the only orthonormal set of functions are the Hermite 
polynomials. They are, however, related by the creation and annihilation operators, that 
is, writing ${\hat a}^\dagger$ for the creation operator we have $H_n(x)=
({\hat a}^\dagger)^n H_0(x)$. The question on the condition required for a set of functions 
$\{\varphi_n(x)\}$, with $\varphi_n(x)=({\hat a}^\dagger)^n \varphi_0(x)$, to form a Reisz 
basis in this Gaussian Hilbert space is of considerable interest, however, because the 
dilation operator does not play a role here, there seems to be no obvious connection to 
the theory of zeta functions. 

\vspace{0.25cm} 
\noindent 
{\bf Acknowledgement}.
DCB thanks the Russian Science Foundation for support (project 16-11-10218).

\end{document}